\documentclass[conference,twocolumn]{IEEEtran}
\IEEEoverridecommandlockouts
\usepackage{cite}
\usepackage{amsmath,amssymb,amsfonts}
\usepackage{graphicx}
\usepackage{tikz}
\usetikzlibrary{IBG}
\usepackage{pgfplots}
\usetikzlibrary{plotmarks}
\usetikzlibrary{arrows}
\usetikzlibrary{spy,backgrounds}
\usepackage{tikz-3dplot}
\usetikzlibrary{positioning,chains,fit,shapes,calc}
\usetikzlibrary{pgfplots.groupplots}
\usepackage{forloop}
\usepackage{tkz-tab}
\usepackage{diagbox}
\usepackage{subcaption}
\pgfplotsset{compat=newest}

\newcommand{\dashedTri}{%
        \begin{tikzpicture}[inner sep=0pt, baseline=(base)]%
        \node[semithick, mark size=2pt,color=black] at (0,0){%
            \pgfuseplotmark{triangle*}%
        };
        \node (base) at (0,-.5ex) {};
        \end{tikzpicture}%
    }

\newcommand{\dashedcross}{%
    \begin{tikzpicture}[inner sep=0pt,baseline=(base)]%
    \node[semithick, mark size=2pt,color=black] at (0,0){%
        \pgfuseplotmark{diamond*}%
    };
    \node (base) at (0,-.5ex) {};
    \end{tikzpicture}%
}
\newcommand{\dashedSquare}{%
    \begin{tikzpicture}[inner sep=0pt,baseline=(base)]%
    \node[semithick, mark size=2pt,color=black] at (0,0){
        \pgfuseplotmark{square*}%
    };
    \node (base) at (0,-.5ex) {};
    \end{tikzpicture}%
}

\newcommand{\dashedOMark}{%
    \begin{tikzpicture}[inner sep=0pt,baseline=(base)]%
    \node[semithick, mark size=2pt,color=black] at (0,0){%
        \pgfuseplotmark{otimes*}%
    };
    \node (base) at (0,-.5ex) {};
    \end{tikzpicture}%
}

\definecolor{DarkBlue}{rgb}{0, 0.2706, 0.541}
\definecolor{BPColor}{rgb}{0.92, 0.43, 0}
\definecolor{IBColor}{rgb}{0.098, 0.4784, 0.5176}
\definecolor{MinSumColor}{rgb}{0.7529, 0, 0}
\definecolor{MinSumColor2}{rgb}{0.498, 0.498, 0.498}

\definecolor{R05Color}{rgb}{0, 0, 0}
\definecolor{R13color}{rgb}{0.098, 0.4784, 0.5176}
\definecolor{R23color}{rgb}{0.7529, 0, 0}

\renewcommand{\vec}[1]{\mathbf{#1}}

\newcommand{\yv}{\vec{y}}

\newcommand{\tch}{t_{ch}}
\newcommand{\Tch}{\mathcal{T}_{ch}}
\newcommand{\Teventch}{|\mathcal{T}_{ch}|}

\newcommand{\tp}{^{\text{T}}}

\begin{document}
\bstctlcite{IEEEexample:BSTcontrol}

%

	\title{Information Bottleneck Decoding of Rate-Compatible 5G-LDPC Codes}
	\author {
		\makebox[.5\linewidth]{Maximilian Stark, Gerhard Bauch}\\ \textit{Hamburg University of Technology} \\	\textit{Institute of Communications}\\\{maximilian.stark,bauch\}@tuhh.de\\
		\and \makebox[.5\linewidth]{Linfang Wang, Richard D. Wesel}\\\textit{University of California, Los Angeles}\\	\textit{Department of Electrical and Computer Engineering}\\					\{lfwang,wesel\}@ucla.edu%
		\thanks{This work has been submitted for publication at the IEEE International Conference on Communications (ICC'20). Copyright may be transferred without notice, after which this version may no longer be accessible.}
	}

	\maketitle

	\begin{abstract}
The new 5G communications standard increases data rates and supports low-latency communication that places constraints on the computational complexity of channel decoders. 5G low-density parity-check (LDPC) codes have the so-called protograph-based raptor-like (PBRL) structure which offers inherent rate-compatibility and excellent performance.  Practical LDPC decoder implementations use message-passing decoding with finite precision, which becomes coarse as complexity is more severely constrained.  Performance degrades as the precision becomes more coarse. Recently, the information bottleneck (IB) method was used to design mutual-information-maximizing lookup tables that replace conventional finite-precision node computations. The IB approach exchanges messages represented by integers with very small bit width. This paper extends the IB principle to the flexible class of PBRL LDPC codes as standardized in 5G. The extensions include puncturing and rate-compatible IB decoder design. As an example of the new approach, a 4-bit information bottleneck decoder is evaluated for PBRL LDPC codes over a typical range of rates.  Frame error rate simulations show that the proposed scheme outperforms offset min-sum decoding algorithms and operates very close to 
double-precision sum-product belief propagation decoding.
\end{abstract}

\section{Introduction}
Low-density parity-check (LDPC) codes are used in the current 5G standard based on their powerful error-correction performance \cite{Richardson18}. The theoretically achievable performance under message passing decoding requires high-precision message representations and computationally complex node operations. Such implementations introduce impractical latency and high power consumption. The desired throughput and latency promised by 5G \cite{Richardson18} require practical hardware implementations that use finite-precision message passing algorithms and node computations that are simplified by smart approximations. Still, the error-rate performance of such finite-precision decoders deteriorates significantly with decreasing precision \cite{Meidlinger.2017}.

To address the challenge of good performance with low precision and simple computation, the \textit{information bottleneck} (IB) decoder \cite{Lewandowsky15,Lewandowsky.2018,applsci-355896,Romero.2016,Meidlinger.2017}
combines ideas from information theory and machine learning.  IB decoders differ from conventional finite-precision decoders significantly. 
First, instead of executing any conventional arithmetic exactly or approximated, the node operations are replaced by \textit{relevant-information-maximizing} functions which map discrete input messages onto discrete output messages.

While similar in operation to the look-up tables developed for finite-alphabet iterative decoding (FAID) approach \cite{Declercq13,Planjery13}, the tables used in IB decoders are learned in an unsupervised manner, using the IB method \cite{Lewandowsky15,Lewandowsky.2018,applsci-355896,Romero.2016,Meidlinger.2017}.

As with the FAID approach, in the entire decoder no log-likelihood ratios (LLRs) are processed at any time. Instead, integer-valued messages, sometimes called \textit{cluster indices}, are exchanged. However, whereas the FAID approach is mainly restricted to regular LDPC codes with variable node degree three,  in our previous work \cite{Lewandowsky.2016,Lewandowsky.2018,Bauch.2018}, IB decoders with only 4 bits of precision perform within $0.1$ dB of double precision belief-propagation for arbitrary regular and arbitrary irregular LDPC codes without puncturing.  For irregular codes, message alignment provides a common representation across nodes with different degrees \cite{applsci-355896}.
In \cite{Ghanaatian.2017} it was shown that, with a similar decoder, a throughput up to 500 Gb/s is possible with high energy and area efficiency.

To the best of our knowledge, all information bottleneck decoders in literature are tailored for 
a specific rate. However, in practical systems a rate-compatible decoding scheme is favorable.
Recently, so-called protograph-based raptor-like (PBRL) LDPC codes were shown to pair very powerful error-correcting capabilities and an efficient structure which enables an inherent rate-compatibility \cite{Ranganathan19}.

This paper presents a generalized design of IB decoders to enable decoding of 5G-related PBRL codes with a bit-width down to 4 bits while incorporating puncturing and hence rate-compatibility into the IB decoder itself.
	In detail, the paper contains the following main contributions:
	\begin{itemize}
		\item This paper extends the design of IB LDPC decoders from \cite{Lewandowsky.2018, applsci-355896} to include puncturing in both the high-rate mother code and the degree-one variable nodes of PBRL codes.
		\item This paper reframes message alignment as its own IB problem, facilitating designs for irregular LDPC codes.
		\item Using the new approach,  a 4-bit information bottleneck decoder for a PBRL code family outperforms a 6-bit offset-min-sum decoder and performs within 0.2 dB of double precision belief propagation decoding.
	\end{itemize}
	\subsubsection*{Organisation}
	The IB method and PBRL LDPC codes are briefly reviewed in Section \ref{seq:prerequisites}.
	In Section \ref{sec:IBdecoders}, we summarize the design of IB LDPC decoders. Thereafter, we use \textit{message alignment} to incorporate puncturing. Finally, this paper targets the problem of rate-compatible decoding architectures in Section \ref{sec:rate_comp_design}. In Section \ref{sec:results}, numerical simulations comparing the performance of our proposed decoder with several reference systems are provided.
	Section \ref{sec:conclusion} concludes the paper.

		\subsubsection*{Notation}
	The realizations $y \in \mathcal{Y}$ from the event space $\mathcal{Y}$ of a discrete random variable $Y$ occur with probability $\Pr(Y=y)$ and $p(y)$ is the corresponding probability distribution. The cardinality or alphabet size of a random variable is denoted by $|\mathcal{Y}|$. Joint distributions and conditional distributions are denoted $p(x,y)$ and $p(x|y)$.

	\section{Prerequisites}
	\label{seq:prerequisites}

	This section briefly reviews the information bottleneck method and its applications in signal processing. Furthermore binary protograph-based raptor-like (PBRL) LDPC codes are introduced.

\subsection{The Information Bottleneck Method}

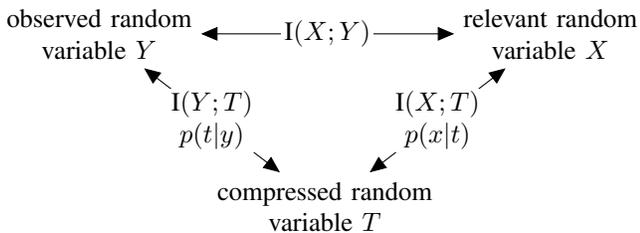
\begin{figure}
	\centering
	\begin{tikzpicture}
	\node (A) at (-3,0) [align=center]{observed random \\ variable $Y$};
	\node (B) at (3,0) [align=center]{relevant random \\ variable $X$ };
	\node (C) at (0,-2.3) [align=center]{compressed random \\ variable $T$ };
	\draw[<->] (A) -- (B) node[fill=white,inner sep=1pt,midway] {I$(X;Y)$};
	\draw[<->] (A) -- (C) node[fill=white,inner sep=1pt,midway,align=center] {I$(Y;T)$ \\ $p(t|y)$ };
	\draw[<->] (C) -- (B) node[fill=white,inner sep=1pt,midway,align=center] {I$(X;T)$ \\ $p(x|t)$} ;

	\end{tikzpicture}
	\caption{Information Bottleneck setup, where I$(X;T)$ is the relevant information, I$(X;Y)$ is the original mutual information and I$(Y;T)$ is the compression information.}
	\vskip -15pt
	\label{fig:IB_setup}
\end{figure}

The information bottleneck method \cite{Tishby.1999} is a mutual-information-maximizing clustering framework from machine learning. The overall information bottleneck setup is depicted in Figure \ref{fig:IB_setup}.
	It considers a Markov chain $X\rightarrow Y \rightarrow T$ of three random variables.
	$X$ is termed the relevant variable, $Y$ is termed the observation and $T$ is a compressed representation of $Y$. The compression is described by the conditional distribution $p(t|y)$. This compression mapping is designed such that the mutual information I$(X;T)$ is maximized while at the same time the mutual information I$(Y;T)$ is minimized. If the
	mapping $p(t|y)$ uniquely assigns a $t$ to each $y$ with probability $1$, this mapping can be implemented in a lookup table such that $t=f(y)$. Algorithms to find suitable compression mappings are described in \cite{Slonim.2002}. These algorithms require the joint distribution $p(x,y)$ and the desired cardinality $|\mathcal{T}|$ of the compression variable $T$ as inputs. 


\subsection{Protograph-Based Raptor-Like (PBRL) LDPC Codes}

Thorpe \cite{thorpe2003low,divsalar2009capacity} introduced LDPC codes constructed from a protograph, which is a small Tanner graph that describes the connectivity of the overall LDPC Tanner graph. A copy and permute operation referred to as \textit{lifting} obtains the full LDPC parity check matrix from the protograph.

\begin{figure}[tb!]
	\centering
	\includegraphics[scale=0.7]{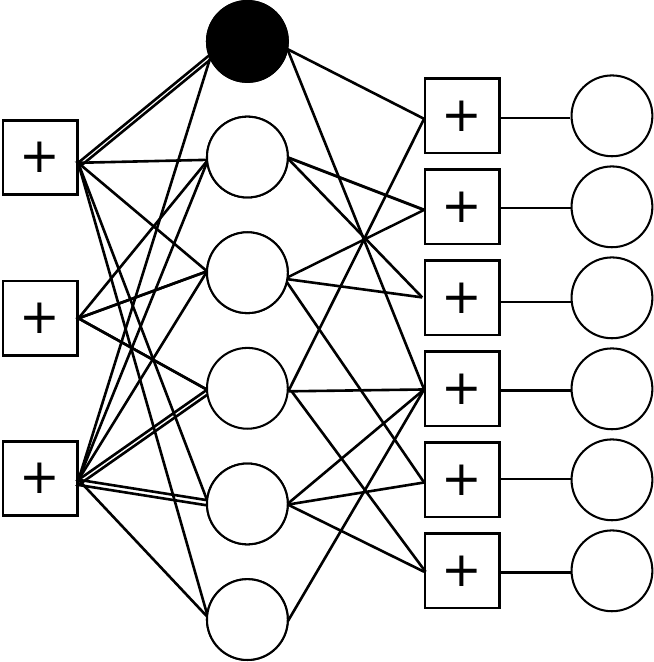}
	\caption{Protograph of a PBRL LDPC code.
	}
	\label{fig:TannerPBRL}
	\vspace{-17pt}
\end{figure}

Figure \ref{fig:TannerPBRL} shows the protograph structure of a PBRL code as described in  \cite{Chen15,Ranganathan19}.
The protograph of an PBRL LDPC Code consists of two parts: (1) a highest-rate code (HRC) protograph and (2) an incremental redundancy code (IRC) protograph. Thereby, the IRC provides lower rates as more of its variable nodes are transmitted, starting from the top. For a more detailed introduction to PBRL LDPC codes we refer the reader to \cite{Chen15,Ranganathan19}.

This paper addresses the issue of designing IB decoders that accommodate the puncturing that is inherent to PBRL code families.  As pointed out in \cite{Chen15}, one or two variable nodes in the HRC are typically punctured, as indicated by the shaded HRC variable node in Figure \ref{fig:TannerPBRL}.  Thus, the IB decoder for the HRC must be designed to handle this puncturing. 
Additionally, all of the IRC variable nodes are punctured for the HRC, but degree-one variable nodes are added to the protograph as the rate is lowered.  The IB decoder must be able to adapt to handle the induced changes in the degree distributions and the associated changes in the probability distributions of message reliabilities that occur as the rate is lowered.

%

\section{Information Bottleneck Decoding of LDPC Codes}
\label{sec:IBdecoders}

In the following section, this paper introduces all the required steps to construct an information bottleneck decoder as described in  \cite{Lewandowsky.2018} and \cite{applsci-355896}.

\subsection{Transmission Scheme and Channel Output Quantization}
\label{sec:qtzr}

\begin{figure}
 \centering
 \input{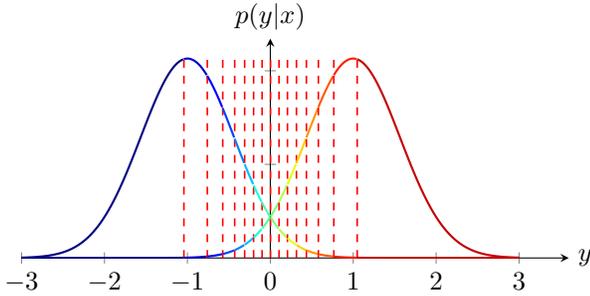}
 \vskip -8pt
 \caption{Quantization boundaries for the BI-AWGN channel computed using the information bottleneck algorithm from \cite{Lewandowsky.2018}. }
\label{fig:AWGN_quanti}
 \vskip -13pt
\end{figure}

We consider binary LDPC codewords transmitted over a symmetric additive white Gaussian noise (AWGN) channel with binary phase shift keying modulation (BPSK) and quantized channel outputs. We use $x$ to denote the equally likely transmitted symbols, which serve as channel inputs.
The binary channel input $x$ and continuous channel output $y$ are related  by the transition probability $p(y|x)$. 
Feeding $p(y,x)$ into the information bottleneck algorithm yields the quantizer mapping $p(\tch|y)$, where  $\tch \in \Tch$ denotes the discrete channel output. Such a mapping is illustrated in Figure \ref{fig:AWGN_quanti}.
In general, a representative log-likelihood ratio can be assigned to each quantization region. These representatives correspond to the quantized channel knowledge which serves as input for sum-product decoding.

In contrast, an information bottleneck decoder does not use any quantized LLRs , but processes a single quantization index $\tch\in\{0,1,\ldots,\Teventch-1\}$ instead.

\subsection{Information Bottleneck Decoders for \textit{Unpunctured} Binary LDPC Codes}

In recent work,
\cite{Lewandowsky15,Lewandowsky.2018,applsci-355896,Romero.2016,Meidlinger.2017}
information bottleneck (IB) decoders were shown to handle the trade-off between low implementation complexity and near-optimal performance very well.
When constructing an IB decoder, node operations are optimized specifically for the available discrete input and output alphabets rather than as approximations of the original belief propagation operations that assume the uncountable  alphabet of all reals for inputs and outputs. The IB operations are essentially look-up tables mapping each possible set of incoming, discrete messages onto a discrete outgoing message. As a result, only highly informative integer-valued messages are passed along the edges of a Tanner graph.

To construct the discrete node operations, the joint probability distribution of the observed random variable $Y$ and relevant random variable $X$ are required (cf. Figure \ref{fig:IB_setup}). In the context of LDPC decoder design, the observed random variables are the $M$ incoming discrete messages $\mathbf{y} = [y_1, \dots, y_M]^{\text{T}}$ and the relevant random variable $X$ depends on the node type. For a variable node, $X$ represents the underlying code bit of a particular node, whereas, if the mapping is designed for a check node, $X$ represents the (mod 2)-sum of the connected, possibly different code bits $b_1, \dots, b_M$. 
Given the joint distribution $p(x,\mathbf{y})$ at each node type in every iteration, the information bottleneck method allows to squeeze $p(x,\mathbf{y})$ through a compact bottleneck. Meaning that the high-dimensional discrete observation vector $\mathbf{y}$, is mapped onto a a scalar integer-valued cluster index $t\in \mathcal{T} =\lbrace 0,1, \dots, |\mathcal{T}|-1 \rbrace $ defined by the mapping $p(t|\mathbf{y})$. 
However, the information bottleneck method aims to preserve all relevant information such that $\text{I}(X;T) \approx \text{I}(X;\mathbf{Y})$. Hence, at any time, $X$ can be inferred very precisely using $p(x|t)$ (cf. Figure \ref{fig:IB_setup}).
However, once the mappings are found, the actual decoding simplifies to look-ups in offline generated tables, which map the sequence of incoming integers $\mathbf{y}$ onto an outgoing integer-valued message $t$. Therefore, instead of passing the meaning $p(x|t)$, e.g. as LLR, only cluster indices are passed which are never converted to an explicit LLR representation.

\subsection{Message Alignment}

\begin{figure}
	\centering
	\begin{tikzpicture}
	\node (A) at (-3,0) [align=center]{observed random \\ variable $(T,D)$};
	\node (B) at (3,0) [align=center]{relevant random \\ variable $X$ };
	\node (C) at (0,-2.6) [align=center]{compressed random \\ variable $Z$ };
	\draw[<->] (A) -- (B) node[fill=white,inner sep=1pt,midway] {$\text{I}(X;T,D)$};
	\draw[<->] (A) -- (C) node[fill=white,inner sep=1pt,midway,align=center] {$\text{I}(T,D;Z)$ \\ $p(z|t,d)$ };
	\draw[<->] (C) -- (B) node[fill=white,inner sep=1pt,midway,align=center] {$\text{I}(X;Z)$ \\ $p(x|z)$} ;

	\end{tikzpicture}
	\caption{The message alignment problem posed as an Information Bottleneck, where $\text{I}(X;Z)$ is the relevant information 
	and $\text{I}(Y;T,D)$ is the compressed information.}
	\vskip -15pt
	\label{fig:MA_setup}
\end{figure}
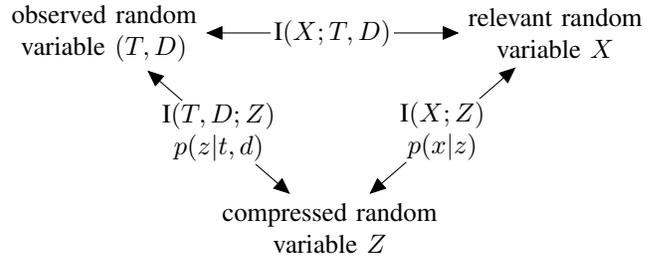

In contrast to regular LDPC codes, irregular LDPC codes are characterized by nodes with varying degrees, i.e., the number of incoming messages differs. This paper leverages the
edge-degree distribution:
\begin{equation}
\lambda(z) =  \sum_{d = 2 }^{\lambda_{\text{max}}} \lambda_d z^{d-1}  \hspace{2cm} \rho(z)  = \sum_{d=2}^{\rho_{\text{max}}}  \rho_d z^{d-1} ,
\end{equation}
where $\lambda_d$ denotes the fraction of edges connected to variable nodes with degree $d$ and  $\rho_d$ denotes the fraction of edges connected to check nodes with degree $d$.
Information bottleneck decoders pass integer-valued cluster indices $t$ instead of log-likelihood ratios.
As the eventspace of $\mathcal{T}$ is \textit{independent} of the node degree $d$ but in contrast the particular meaning $p(x|t,d)$ \textit{depends} on the node degree, a receiving node cannot resolve if the message originates from a conveying with high or low degree from the cluster indices alone. As the variety of node degrees increases also the range of the different meaning increases likewise. In turn, the decoder cannot exploit the full error-correction capabilities of the LDPC code because it cannot uniquely recover the actual meaning of the received message.
In \cite{applsci-355896} it was revealed that with a processing at the origin nodes called \textit{message alignment}, information bottleneck decoders can achieve competitive performance for LDPC codes with non-optimized irregular degree distributions. As an extension to the original mutual-information based look-up table design, message alignment considers the three random variables $T,X$ and $D$ jointly, i.e, the clusters, the channel input (which is the relevant random variable), \textit{and} the node degree. Given the node dependent meanings $p(x,t|d)$ and the edge degree distributions, the task is to find a mapping $p(z|t,d)$ such that $\text{I}(X;T,D) \approx \text{I}(X;Z)$ \cite{applsci-355896}. 
Using this modified objective, message alignment ensures that identical outgoing cluster indices of nodes with different degrees correspond to similar beliefs on the code bits. 
In fact, Figure \ref{fig:MA_setup} reveals that the objective of message alignment is closely related to the information bottleneck setup from Figure \ref{fig:IB_setup}. Leveraging the information bottleneck notation, one would refer to $X$ as the relevant quantity and $T$ \textit{and} $D$ represent the observation which squeeze through a compact bottleneck resulting in $Z$ which is a compressed version of $T$ \textit{and} $D$ but preserves the maximum relevant information $\text{I}(X;Z)$.

\section{Information Bottleneck PBRL LDPC Decoders}
\label{sec:rate_comp_design}
To decode PBRL LDPC codes the respective decoders must support puncturing. Although puncturing itself is a fairly easy problem for conventional decoders, the design of information bottleneck decoders prohibits the use of puncturing. This section contains our main contributions. First, this paper shows how to incorporate punctured nodes using message alignment. Second, this paper devises a generalized scheme well suited for the structure of PBRL codes enabling rate-compatibility.

\subsection{Constructing Information Bottleneck Decoders for Punctured PBRL LDPC Codes }

To increase the code rate, ``punctured'' codeword bits are not transmitted and are thus unknown to the receiver. In conventional approaches this corresponds to an LLR=0 for the respective punctured bits.

\begin{figure}
 \centering
 \begin{tikzpicture}[hv path/.style={to path={-| (\tikztotarget)}},
vh path/.style={to path={|- (\tikztotarget)}}]

\matrix [row sep=2mm,column sep=2mm] {
	\node[latent]	(y0) {$t_\text{ch}^\text{in}$};& & & &; \\
	\node[latent]	(y1) {$t_1^\text{in}$}; &; \IBGnode {IBN_21} {$x$} {} {}; & \node[latent] (t21) {$t_1$};& & &  &  &[10mm] &\node[latent] (z21) {$z_1$}; \\
	\node[latent] (y2) {$t_2^\text{in}$};& & & \IBGnode {IBN_31} {$x$} {} {}; & \node[latent] (t31) {$t_2$}; & & & &\node[latent] (z31) {$z_2$}; \\
    \node[latent] (y3) {$t_3^\text{in}$};& & & & & \IBGnode {IBN_41} {$x$} {} {}; & \node[latent] (t41) {$t_3$};& &\node[latent] (z41) {$z_3$}; \\
%
};


\graph [use existing nodes] {
	y0 -- IBN_21 --t21 -- IBN_31;
	t21 --[dotted] z21;
	y1 -- IBN_21;
	y2 -- IBN_31 -- t31 -- IBN_41;
	t31 --[dotted] z31;
	y3 -- IBN_41 -- t41;
	t41 --[dotted] z41;
	
	};

\node[right= 28pt of y0.north,inner sep=0pt] (circ1) {$d_v=2$};
\node[draw, rectangle,dashdotted, fit=(y0)(y1)(IBN_21)(circ1), inner sep=4pt] (plate) {};

\node[right= 75pt of y0.north,inner sep=0pt] (circ2) {$d_v=3$};
\node[draw, rectangle,dashdotted, fit=(y0)(y1)(IBN_31)(circ2), inner sep=4pt] (plate2) {};

\node[right= 123pt of y0.north,inner sep=0pt] (circ3) {$d_v=4$};
\node[draw, rectangle,dashdotted, fit=(y0)(y1)(y2)(IBN_41)(circ3), inner sep=4pt] (plate3) {};
\draw (2.2,-2.0) rectangle (2.9,1.0) node[pos=.5] {\rotatebox{-90}{Message Alignment}};



\end{tikzpicture}
 \vskip -2pt
 \caption{Information bottleneck graph for a concatenated lookup table for $d_\text{v}=4$ with message alignment.}
 \label{fig:var}
 \vskip -12pt
\end{figure}
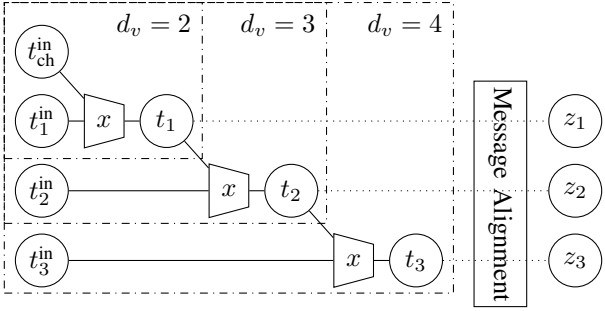

Going back to Figure \ref{fig:AWGN_quanti} reveals that the information-optimum channel quantizer is fully symmetric \cite{Kurkoski.2008}. Typically, the resolution of the analog-to-digital converter (ADC) is a power of two and thus even. In this case, the LLR=0 can not be represented properly. Usually the quantization boundaries are then shifted to be able to represent LLR=0.

In information bottleneck decoders, LLR representations are ignored. Furthermore, the decoder is designed using density evolution and thus requires the complete knowledge of the statistics of the decoder input. Hence, constructing an information bottleneck decoder for puncturing is far less straightforward than in conventional decoders. In the following we propose a generic extension to include puncturing which leverages the message alignment technique.

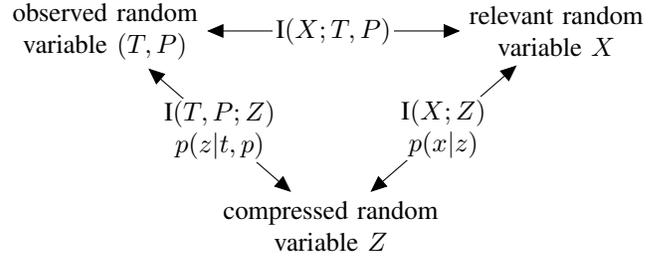
\begin{figure}
	\centering
	\begin{tikzpicture}
	\node (A) at (-3,0) [align=center]{observed random \\ variable $(T,P)$};
	\node (B) at (3,0) [align=center]{relevant random \\ variable $X$ };
	\node (C) at (0,-2.6) [align=center]{compressed random \\ variable $Z$ };
	\draw[<->] (A) -- (B) node[fill=white,inner sep=1pt,midway] {$\text{I}(X;T,P)$};
	\draw[<->] (A) -- (C) node[fill=white,inner sep=1pt,midway,align=center] {$\text{I}(T,P;Z)$ \\ $p(z|t,p)$ };
	\draw[<->] (C) -- (B) node[fill=white,inner sep=1pt,midway,align=center] {$\text{I}(X;Z)$ \\ $p(x|z)$} ;

	\end{tikzpicture}
	\caption{Considering puncturing as message alignment problem, where $\text{I}(X;Z)$ is the relevant information 
	and $\text{I}(Z;T,P)$ is the compressed information.}
	\vskip -12pt
	\label{fig:MA_setup_punct}
\end{figure}

Without loss of generality, we consider a variable node with degree $d_v=4$ which processes one channel message and three messages received from connected check nodes to generate extrinsic information about the underlying code bit. Figure \ref{fig:var} shows this processing as a concatenation of two-input-look-up tables. 
In Figure \ref{fig:var} each look-up table is depicted as trapezoid with the input vector  $\yv=[t^{in}_{ch},t^{in}_1]\tp$ or $\yv=[t_{i-1},t^{in}_i]\tp$ and output $t_i$. The respective mappings are found using the information bottleneck method. 
The joint distribution in the first stage, which is processing the input vector $\yv=[t^{in}_{ch},t^{in}_1]\tp$ is computed as follows (See \cite{Lewandowsky.2018} for details.):
\begin{equation}
  \label{eq:joint_var_node}
  p(x,[y^{in}_{ch},y^{in}_1]\tp) = \frac{1}{p(x)} p(x,y^{in}_{ch}) p(x,y^{in}_1).
\end{equation}
Clearly, $p(x,[t^{in}_{ch},t^{in}_1]\tp)$ and thus also $p(x|t_1)$, which is used in the next step, depend on the statistics of the quantized channel output (cf. Section \ref{sec:qtzr}). When incorporating puncturing, $p(x,t^{in}_{ch})$ differs depending on whether the variable node is punctured or not.
First, we introduce the random variable $P$ with eventspace \{true,false\} indicating if a node is punctured or not and  $\Pr(P=true)=1-\Pr(P=false)$ where $\Pr(P=true)$ is the puncturing rate.
In this paper, the puncturing rate indicates the fraction of variable nodes with degree $d>1$ that are punctured. In the next section, it will become clear why punctured variable nodes with degree $d=1$ are ignored when computing the puncturing rate. Please note that, if the node is punctured, $p(x,t^{in}_{ch}|P=true)$, is uniformly distributed.
As a result, we rewrite \eqref{eq:joint_var_node} as
\begin{equation}
  \label{eq:joint_var_node_new}
  p(x,[y^{in}_{ch},y^{in}_1]\tp|p) = \frac{1}{p(x)} p(x,y^{in}_{ch}|p) p(x,y^{in}_1).
\end{equation}
Due to the concatenation of lookup tables as shown in Figure \ref{fig:var} all subsequent tables depend on $P$. 
Consequently, in a straightforward implementation the number of required lookup tables will increase drastically to account for all possible combinations of punctured and non-punctured nodes and there respective degrees.
This paper proposes a novel design objective which includes puncturing in the table design. In turn, $P$ and $T$ are considered jointly, i.e., the objective is to find a mapping $p(z|t,p)$ such that $\text{I}(X;T,P)\approx \text{I}(X;Z)$. As depicted in Figure \ref{fig:MA_setup_punct} the resulting problem is basically a message alignment setup. 
As a result, one creates the mapping $p(z|t,p)$ and the meaning $p(x|z)$ such that all subsequently constructed tables do not depend any longer on the node being punctured or not. This approach significantly reduces the number of distinct lookup tables.

\subsection{Constructing Information Bottleneck Decoders for Rate-Compatible PBRL LDPC Codes}
PBRL codes can operate close to the theoretical limit for a wide range of code rates \cite{Chen15}. The rate can be easily adapted by puncturing or transmitting the degree-one variables (cf. Figure \ref{fig:TannerPBRL}). 
As a punctured degree-one variable node always conveys an LLR equal to zero to the connected check nodes, these check nodes and thus 
entire parts of the respective Tanner graph are deactivated \cite{Chen15}, meaning that no relevant information propagates from the punctured degree-one nodes towards the inner variable nodes with very high degree (cf. Figure \ref{fig:TannerPBRL}).

Hence, puncturing degree-one nodes, i.e., deactivating parts of the Tanner graph changes the \textit{effective} degree distribution $\lambda_{eff}(z) \neq \lambda(z)$ and $\rho_{eff}(z) \neq \rho(z)$. To allow information bottleneck decoders to cope with punctured degree-one nodes and rate-adaptability the effective degree distribution is of crucial importance. To determine the effective degree distribution, the following strategy is proposed:
\subsubsection*{Variable Node} If an edge carries no information, this would correspond to LLR = 0 in a conventional decoder, the effective node degree seen by the outgoing message is reduced by one.
\subsubsection*{Check Node} If already one edge which is used to generate extrinsic information contributes no information, i.e., the corresponding LLR = 0, the outgoing message will automatically carry no extrinsic information. Thus, only the fraction of edges over which extrinsic information is passed is considered for each effective degree.

The computations are performed offline given a parity check matrix for every iteration.
In Figure \ref{fig:var}, we have depicted an unfold variable node. It can be seen that in the general scheme only one tree of lookup tables for the highest node degree in the code has to be constructed. All nodes with smaller degrees can easily reuse the tables. 
As mentioned above, puncturing degree-one variable nodes reduce the effective degree distribution. However, in the context of the concatenated structure presented in Figure \ref{fig:var}, this paper argues that puncturing basically only effects the depth of the lookup tree, because if a message is punctured it would not contribute any relevant information and thus can be skipped. Hence, designing a decoder for the lowest code rate, yields the maximum depth of the look-up tree. Higher code rates will reuse this architecture but require individually optimized tables. It is possible to reuse the same tables for a range of code rates, but this analysis is left to the journal version of this paper due to space limitations.



\section{Results and Discussion}
\label{sec:results}

In this section, we present and discuss results obtained performing frame error rate simulations for an exemplary PBRL LDPC code. The code was taken from \cite{Chen15}. The code has $K=1032$ information bits, and is evaluated for various code rates $R_\text{c}$ range from $R_\text{c}=1/3$ up to $R_\text{c} = 2/3$.

We propose to construct all involved lookup tables just once for a fixed design-$E_b/N_0$. The constructed lookup tables are then stored and applied for all $E_b/N_0$. Hence, the lookup table construction needs to be done only once and offline.

\begin{table*}[t]
\vskip 5pt
\caption{Simulation parameters}
\vskip-5pt
\centering
\begin{tabular}{lcccccc}
decoder&node operation (check / var) & precision exchanged messages & precision check node & precision variable node & channel quantizer\\
\hline\hline
sum-product& box-plus / addition & 64 bit & 64 bit & 64 bit & None\\
\hline
offset min-sum& $\min^*()$ / addition & 4 bit & 4 bit & 6 bit & 4 bit \\
\hline
layered NMSA& $\min^*()$ / addition & 6 bit & 6 bit & 6 bit & None \\
\hline
proposed& lookup table / lookup table& 4 bit & 4 bit & 4 bit & 4 bit\\
\end{tabular}
\label{tab:params1}
\vskip-12pt
\end{table*}

We consider three reference schemes to compare the performance of our decoder. Decoding of a codeword is stopped after a maximum number of 100 decoding iterations or earlier if the syndrome check is successful. First, we consider a double-precision belief propagation decoder with flooding schedule. The received samples are \textit{not} quantized and the internal operations are additions at the variable node and box-plus at the check node. 
Second, we use the layered normalized min-sum algorithm (NMSA) \cite{chen2002density,zhang2002shuffled} with 6 bit resolution at the check node and 6 bits at the variable node. Again the inputs to the decoder are \textit{not} quantized. 
The operations here are again additions at the variable nodes but the normalized min-sum approximation is used at the check nodes.
Third, we use the offset-min-sum decoder with only 4 bit resolution at the check node and 6 bits at the variable node to prevent an overflow when adding the 4 bit messages received from the channel quantizer. Finally, we designed our proposed information bottleneck decoder for fully 4 bit integer architecture. This means, starting form the channel quantizer which outputs 4 bit integers, the internal messages require only 4 bits and only \textit{lookup} operations performed. These lookups do not mimic any arithmetic function but realize the relevant-information preserving mappings found using the information bottleneck method. 



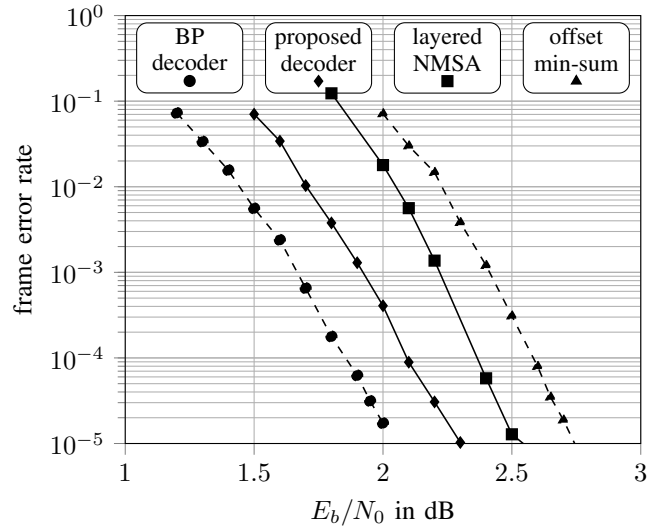
\begin{figure}
	\centering
\begin{tikzpicture}

\begin{axis}[
xmin=1.0, xmax=3,
ymin=1.e-05, ymax=1,
ymode=log,
tick align=outside,
tick pos=left,
xmajorgrids,
xminorgrids,
x grid style={white!69.01960784313725!black},
ymajorgrids,
yminorgrids,
y grid style={white!69.01960784313725!black},
legend pos = north east,
legend cell align={left},
ylabel= frame error rate,
xlabel=$E_b/N_0$ in dB,
]

\addplot [semithick, R05Color,dashed,mark = otimes*] 
table {%
1.2 0.0726666666666667
1.3 0.0336666666666667
1.4 0.015625
1.5 0.00558659217877095
1.6 0.00239234449760766
1.7 0.000652741514360313
1.8 0.000177777777777778
1.9 6.23402531014276e-05
1.95 3.14238129654652e-05
2 1.72699640784747e-05
};


\addplot [semithick, R05Color,dashed,mark=triangle*]
table {%
2 0.07125
2.1 0.03
2.2 0.0147058823529412
2.3 0.00381679389312977
2.4 0.00120192307692308
2.5 0.000308832612723904
2.6 7.95165394402036e-05
2.65 3.48213663904172e-05
2.7 1.8807598269701e-05
2.75 9.11477322444218e-06
};

\addplot [semithick, R05Color,mark=diamond*]
table {%
1.5 0.0704545454545455
1.6 0.0341296928327645
1.7 0.0103199174406605
1.8 0.00378036688150719
1.9 0.00129477772982305
2 0.000406724511930586
2.1 8.9142979396086e-05
2.2 3.06030015423913e-05
2.3 1.02749437960574e-05
};

\addplot [semithick, R05Color,mark=square*]
table {%
1.8	0.123456790123457	
2.0	0.0179115171054989
2.1	0.00561987186692148
2.2	0.00136973166956589	
2.4	5.78074779753468e-05
2.5	1.28132125747360e-05
2.6	7.29347575256778e-06
2.7	5.35195728834648e-06
};

\tikzstyle{mybox4} = [draw=black, fill=white, thin,
rectangle,rounded corners, inner sep=2pt, inner ysep=2pt]
\node [mybox4](boxlog) at (axis cs:2.75,3.2e-1){%
	\begin{minipage}{0.07\textwidth}
	\begin{centering}
	\small {\color{black} offset min-sum}\\[-4pt]
	\small {\color{black} \dashedTri }\\
	\end{centering} 
	\end{minipage}
};  

\tikzstyle{mybox5} = [draw=black, fill=white, thin,
rectangle,rounded corners, inner sep=2pt, inner ysep=2pt]
\node [mybox4](boxlog) at (axis cs:2.25,3.2e-1){%
	\begin{minipage}{0.07\textwidth}
	\begin{centering}
	\small {\color{black} layered NMSA}\\[-4pt]
	\small {\color{black} \dashedSquare}\\
	\end{centering} 
	\end{minipage}
};  

\tikzstyle{mybox2} = [draw=black, fill=white, thin,
rectangle,rounded corners, inner sep=2pt, inner ysep=2pt]
\node [mybox2](boxbp) at (axis cs:1.25,3.2e-1){%
	\begin{minipage}{0.07\textwidth}
	\begin{centering}
	\small { \color{black}  BP \\ decoder\\[-4pt] \dashedOMark} \\
	\end{centering}
	\end{minipage}
};  


\tikzstyle{mybox3} = [draw=black, fill=white, thin,
rectangle,rounded corners, inner sep=2pt, inner ysep=2pt]
\node [mybox3](boxib) at (axis cs:1.75,3.2e-1){%
	\begin{minipage}{0.07\textwidth}
	\begin{centering}
	\small {\color{black}proposed decoder \\[-4pt] \dashedcross }\\
	\end{centering}
	\end{minipage}
};  

\end{axis}

\end{tikzpicture}
	\vskip -5pt
	\caption{Frame error rates 
	for the proposed scheme (diamond-marker), and the reference schemes summarized in Table \ref{tab:params1} 
		for the considered PBRL LDPC code with code rate $R_c=1/2$.}
	\label{fig:FER}
	\vskip -10pt
\end{figure}

\begin{figure}
	\centering
\begin{tikzpicture}

\begin{axis}[
xmin=0.4, xmax=3.55,
ymin=1.e-05, ymax=1,
ymode=log,
tick align=outside,
tick pos=left,
xmajorgrids,
xminorgrids,
x grid style={white!69.01960784313725!black},
ymajorgrids,
yminorgrids,
y grid style={white!69.01960784313725!black},
legend style={at={(0.03,0.03)}, anchor=south west, draw=white!80.0!black},
legend cell align={left},
ylabel= frame error rate,
xlabel=$E_b/N_0$ in dB,
]

\addplot [semithick, R13color,dashed,mark = otimes*] 
table {%
0.5 0.0525
0.6 0.0265789473684211
0.7 0.0108695652173913
0.8 0.0045662100456621
0.9 0.00151515151515152
1 0.000605693519079346
1.1 0.000207425845260319
1.2 8.93415527561869e-05
1.25 4.30885901413306e-05
1.3 2.59376458992582e-05
1.35 1.74880207058165e-05
1.4 1.2987012987013e-05
1.45 7.25447238222364e-06
};


\addplot [semithick, R23color,densely dotted,mark = otimes*] 
table {%
2.2 0.0535
2.3 0.0227272727272727
2.4 0.0113483146067416
2.5 0.003125
2.6 0.00115340253748558
2.7 0.000342231348391513
2.8 0.000109277674571085
2.9 2.55004462578095e-05
2.95 1.60642570281125e-05
};


\addplot [semithick, R13color,dashed,mark=triangle*] 
table {%
2 0.0588888888888889
2.1 0.026
2.2 0.00847457627118644
2.3 0.00324675324675325
2.4 0.000978473581213307
2.5 0.000194704049844237
2.6 4.18130122093996e-05
2.65 2.09415312447646e-05
};

\addplot [semithick, R23color,dashed,mark=triangle*] 
table {%
2.8 0.0846153846153846
2.9 0.0416666666666667
3 0.02
3.1 0.00537634408602151
3.2 0.0025706940874036
3.3 0.000578703703703704
3.4 0.000151883353584447
3.5 2.29584222972197e-05
3.55 1.01187946491814e-05
};


\addplot [semithick, R13color,dashed,mark=diamond*]
table {%
1.075 0.0272232304900181
1.17 0.0171821305841924
1.265 0.00602772754671489
1.36 0.00261917234154007
1.455 0.000995784512231553
1.55 0.00044407602581562
1.645 0.000132545717226967
1.74 4.2061814041916e-05
1.835 1.5742255203996e-05
1.93 4.77904484416729e-06
2.025 3.06414162083732e-06
};

\addplot [semithick, R23color,dashed,mark=diamond*]
table {%
2.3 0.0665938864628821
2.4 0.0393184796854522
2.5 0.0222057735011103
2.6 0.00467216944401184
2.7 0.00190282887225675
2.8 0.000730246823426318
2.9 0.000146517284155621
3 3.37306807525989e-05
3.1 8.89312596935073e-06
3.2 4.68597742342937e-06
};

\addplot [semithick, R13color,dashed,mark=square*]
table {%
	1.6 0.0143061516452074
	1.8 0.000819289348419128	
	2.0 3.02233322916345e-05
	2.1 8.58873216746936e-06	
	2.2 5.97369316945873e-06
};

\addplot [semithick, R23color,dashed,mark=square*]
table {%
	2.4 0.0899280575539568
	2.8 0.000889766792123825
	3.0 5.14029932990878e-05	
	3.2 1.83650007162806e-05
	3.3 1.73566032155570e-05
};

\tikzstyle{mybox4} = [draw=black, fill=white, thin,
rectangle,rounded corners, inner sep=2pt, inner ysep=2pt]
\node [mybox4](boxlog) at (axis cs:3.2,3.2e-1){%
	\begin{minipage}{0.07\textwidth}
	\begin{centering}
	\small {\color{black} offset min-sum}\\[-4pt]
	\small {\color{black} \dashedTri }\\
	\end{centering} 
	\end{minipage}
};  

\tikzstyle{mybox5} = [draw=black, fill=white, thin,
rectangle,rounded corners, inner sep=2pt, inner ysep=2pt]
\node [mybox4](boxlog) at (axis cs:2.4,3.2e-1){%
	\begin{minipage}{0.07\textwidth}
	\begin{centering}
	\small {\color{black} layered NMSA}\\[-4pt]
	\small {\color{black} \dashedSquare}\\
	\end{centering} 
	\end{minipage}
};  

\tikzstyle{mybox2} = [draw=black, fill=white, thin,
rectangle,rounded corners, inner sep=2pt, inner ysep=2pt]
\node [mybox2](boxbp) at (axis cs:0.8,3.2e-1){%
	\begin{minipage}{0.07\textwidth}
	\begin{centering}
	\small { \color{black}  BP \\ decoder\\[-4pt] \dashedOMark} \\
	\end{centering}
	\end{minipage}
};  


\tikzstyle{mybox3} = [draw=black, fill=white, thin,
rectangle,rounded corners, inner sep=2pt, inner ysep=2pt]
\node [mybox3](boxib) at (axis cs:1.6,3.2e-1){%
	\begin{minipage}{0.07\textwidth}
	\begin{centering}
	\small {\color{black}proposed decoder \\[-4pt] \dashedcross }\\
	\end{centering}
	\end{minipage}
};

\end{axis}
\def\a{2.0} 
\def\b{0.08} 
\draw (3,1) ellipse ({\a} and {\b});
\node at (0.6,0.6) {$R=\frac{1}{3}$};

\def\a{1.0} 
\def\b{0.08} 
\draw (5.5,2) ellipse ({\a} and {\b});
\node at (6.05,2.5) {$R=\frac{2}{3}$};
\end{tikzpicture}
	\vskip -5pt
	\caption{Frame error rates 
	for the proposed scheme (diamond-marker) and the reference schemes summarized in Table \ref{tab:params1} 
		for the considered PBRL LDPC code with code rate $R_c=1/3$ (blue, dashed),  $2/3$ (red, dotted).}
	\label{fig:FER_rate_compatible}
	\vskip -12pt
\end{figure}
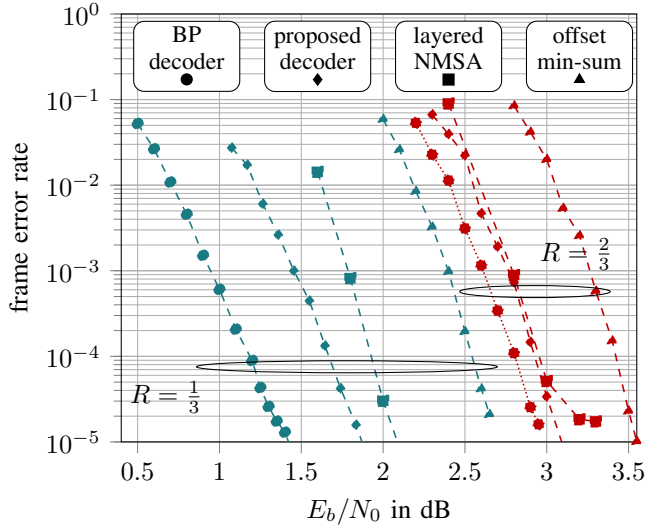

The most important parameters of the applied decoders are again summarized in Table \ref{tab:params1} for a quick overview.
First we consider a decoder designed for a fixed rate of $R= 0.5$. The results are shown in Figure \ref{fig:FER}.
As expected the sum-product algorithm (o-marker) achieves the best frame error rate performance, but at the same time, has the highest computational complexity (cf. Table \ref{tab:params1}). Although all applied operations
in the information bottleneck decoder (x marker) are simple lookups, the decoder performs only less than $0.2$ dB worse than the benchmark. 
The results are even more remarkable when considering the tremendous gap to the two offset-min-sum with an even slightly higher resolution. Please note, that PBRL codes have typically variable nodes with very large degrees. From the gap of 0.75 dB noticed in Figure \ref{fig:FER} we conclude that a conventional offset-min-sum decoder which exchanges only 4 bit messages can not be used for PBRL codes with such a coarse quantization, since the dynamic range of the LLRs cannot be captured appropriately. The gap can be reduced by choosing a finer resolution as indicated by the frame error rate curve for the 6 bit NMSA decoder.
However, with the generalized design for information bottleneck decoders proposed in this paper, both challenges, i.e., puncturing and rate-compatible design can be efficiently tackled to enable fully 4 bit decoders for PBRL codes. Figure \ref{fig:FER_rate_compatible} shows results for various other rates. 
For all considered rates, the belief propagation decoder with double-precision resolution and no channel quantizer achieves the best performance. However, 
again we observe that the proposed information bottleneck decoder operates very close to this benchmark. Interestingly, the proposed schemes outperforms the 4 bit offset min-sum decoder and the 6 bit NMSA decoder for all investigated rates.


\section{Conclusion}
This paper uses the information bottleneck method to efficiently represent reliability information, reducing the data transfer and computational complexity of 5G protograph-based raptor-like LDPC decoding. The proposed decoder extends the information bottleneck method to incorporate puncturing and leverages the inherent rate-compatibility of this powerful class of LDPC codes to develop a rate-compatible decoder. The proposed information bottleneck framework integrates a message alignment module to dynamically adjust to the degree distribution of input messages. This approach accommodates puncturing for all supported rates without significantly increasing the number of required information bottleneck lookup tables. The proposed information bottleneck decoder exchanges only 4 bit integers and replaces the arithmetic in the node operations by lookup tables. This decoder performs only $0.2$ dB worse than the sum-product algorithm and outperforms the offset-min-sum algorithm.  Future work will investigate information bottleneck decoders with flexible bit widths to allow performance to be optimized based on available memory and a table reuse for several code rates.

\label{sec:conclusion}
\bibliographystyle{IEEEtran}
\bibliography{mybib}

\begin{thebibliography}{10}
\providecommand{\url}[1]{#1}
\csname url@samestyle\endcsname
\providecommand{\newblock}{\relax}
\providecommand{\bibinfo}[2]{#2}
\providecommand{\BIBentrySTDinterwordspacing}{\spaceskip=0pt\relax}
\providecommand{\BIBentryALTinterwordstretchfactor}{4}
\providecommand{\BIBentryALTinterwordspacing}{\spaceskip=\fontdimen2\font plus
\BIBentryALTinterwordstretchfactor\fontdimen3\font minus
  \fontdimen4\font\relax}
\providecommand{\BIBforeignlanguage}[2]{{%
\expandafter\ifx\csname l@#1\endcsname\relax
\typeout{** WARNING: IEEEtran.bst: No hyphenation pattern has been}%
\typeout{** loaded for the language `#1'. Using the pattern for}%
\typeout{** the default language instead.}%
\else
\language=\csname l@#1\endcsname
\fi
#2}}
\providecommand{\BIBdecl}{\relax}
\BIBdecl

\bibitem{Richardson18}
T.~{Richardson} and S.~{Kudekar}, ``Design of low-density parity check codes
  for {5G} new radio,'' \emph{IEEE Communications Magazine}, vol.~56, no.~3,
  pp. 28--34, March 2018.

\bibitem{Meidlinger.2017}
M.~Meidlinger and G.~Matz, ``On irregular {LDPC} codes with quantized message
  passing decoding,'' in \emph{Proc. SPAWC'17}.\hskip 1em plus 0.5em minus
  0.4em\relax Piscataway, NJ: IEEE, 2017, pp. 1--5.

\bibitem{Lewandowsky15}
J.~{Lewandowsky} and G.~{Bauch}, ``Trellis based node operations for {LDPC}
  decoders from the information bottleneck method,'' in \emph{Proc. ICSPCS'15},
  Dec 2015, pp. 1--10.

\bibitem{Lewandowsky.2018}
J.~Lewandowsky and G.~Bauch, ``Information-optimum {LDPC} decoders based on the
  information bottleneck method,'' \emph{IEEE Access}, vol.~6, pp. 4054--4071,
  2018.

\bibitem{applsci-355896}
M.~Stark, J.~Lewandowsky, and G.~Bauch, ``Information-bottleneck decoding of
  high-rate irregular {LDPC} codes for optical communication using message
  alignment,'' \emph{Applied Sciences}, vol.~8, no.~10, 2018.

\bibitem{Romero.2016}
F.~J.~C. Romero and B.~M. Kurkoski, ``{LDPC} {decoding} {mappings} {that}
  {maximize} {mutual} {information},'' \emph{IEEE Journal on Selected Areas in
  Communications}, vol.~34, no.~9, pp. 2391--2401, Sep. 2016.

\bibitem{Declercq13}
D.~{Declercq} \emph{et~al.}, ``Finite alphabet iterative decoders—part ii:
  Towards guaranteed error correction of {LDPC} codes via iterative decoder
  diversity,'' \emph{IEEE Transactions on Communications}, vol.~61, no.~10, pp.
  4046--4057, October 2013.

\bibitem{Planjery13}
S.~K. {Planjery} \emph{et~al.}, ``Finite alphabet iterative decoders—part i:
  Decoding beyond belief propagation on the binary symmetric channel,''
  \emph{IEEE Transactions on Communications}, vol.~61, no.~10, pp. 4033--4045,
  October 2013.

\bibitem{Lewandowsky.2016}
J.~Lewandowsky, M.~Stark, and G.~Bauch, ``Information bottleneck graphs for
  receiver design,'' in \emph{Proc. ISIT'16}, Barcelona, Spain, Jul. 2016, pp.
  2888--2892.

\bibitem{Bauch.2018}
G.~Bauch \emph{et~al.}, ``Information-optimum discrete signal processing for
  detection and decoding,'' in \emph{Proc. VTC-Spring'18}, Porto, Portugal,
  2018.

\bibitem{Ghanaatian.2017}
R.~Ghanaatian \emph{et~al.}, ``A 588-gb/s {LDPC} decoder based on
  finite-alphabet message passing,'' \emph{IEEE Transactions on Very Large
  Scale Integration (VLSI) Systems}, vol.~26, no.~2, pp. 329--340, Feb. 2018.

\bibitem{Ranganathan19}
S.~V.~S. {Ranganathan}, D.~{Divsalar}, and R.~D. {Wesel}, ``Quasi-cyclic
  protograph-based raptor-like ldpc codes for short block-lengths,'' \emph{IEEE
  Trans. on Info. Theory}, vol.~65, no.~6, pp. 3758--3777, June 2019.

\bibitem{Tishby.1999}
N.~Tishby, F.~C. Pereira, and W.~Bialek, ``The information bottleneck method,''
  in \emph{Proc. 37th Allerton Conference on Communication and
  Computation}.\hskip 1em plus 0.5em minus 0.4em\relax IEEE, 1999.

\bibitem{Slonim.2002}
N.~Slonim, ``The information bottleneck: Theory and applications,'' Ph.D.
  dissertation, Hebrew University of Jerusalem, 2002.

\bibitem{thorpe2003low}
J.~Thorpe, ``Low-density parity-check ({LDPC}) codes constructed from
  protographs,'' \emph{IPN progress report}, vol.~42, no. 154, pp. 42--154,
  2003.

\bibitem{divsalar2009capacity}
D.~Divsalar \emph{et~al.}, ``Capacity-approaching protograph codes,''
  \emph{IEEE Journal on Selected Areas in Communications}, vol.~27, no.~6, pp.
  876--888, 2009.

\bibitem{Chen15}
T.~{Chen} \emph{et~al.}, ``Protograph-based raptor-like {LDPC} codes,''
  \emph{IEEE Transactions on Communications}, vol.~63, no.~5, pp. 1522--1532,
  May 2015.

\bibitem{Kurkoski.2008}
B.~M. Kurkoski, K.~Yamaguchi, and K.~Kobayashi, ``Noise thresholds for discrete
  {LDPC} decoding mappings,'' in \emph{Proc. {GLOBECOM}'18}, New Orleans, LO,
  USA, Dec. 2008, pp. 1--5.

\bibitem{chen2002density}
J.~Chen, ``Density evolution for {BP}-based decoding algorithms of {LDPC} codes
  and their quantized versions,'' in \emph{Proc. IEEE GLOBECOM'02},
  vol.~2.\hskip 1em plus 0.5em minus 0.4em\relax Taipei, Taiwan: IEEE, 2002,
  pp. 1378--1382.

\bibitem{zhang2002shuffled}
J.~Zhang and M.~Fossorier, ``Shuffled belief propagation decoding,'' in
  \emph{Proc. 36th Asilomar Conference on Signals, Systems and Computers},
  vol.~1.\hskip 1em plus 0.5em minus 0.4em\relax IEEE, 2002, pp. 8--15.

\end{thebibliography}

\end{document}